\documentstyle{article}

\begin{document}
\begin{center}
{\Large\bf
Bremsstrahlung of relativistic electrons \\
in the Aharonov-Bohm potential
}\\[1cm]
{\large
J\"urgen Audretsch\footnote{e-mail:
Juergen.Audretsch@uni-konstanz.de} and
Ulf Jasper\footnote{e-mail:
Ulf.Jasper@uni-konstanz.de}}\\
\medskip
Fakult\"at f\"ur Physik der Universit\"at Konstanz,
Postfach 5560, D 78434 Konstanz, Germany
\\[0.5cm]
{\large Vladimir D.~Skarzhinsky\footnote{e-mail:
vdskarzh@sgi.lpi.msk.su}\\
}
\medskip
Fakult\"at f\"ur Physik der Universit\"at Konstanz,
Postfach 5560, D 78434 Konstanz, Germany \\ and
P.~N.~Lebedev Physical Institute,
Leninsky prospect 53, Moscow 117924, Russia

\bigskip
(Phys.Rev.,D, {\bf 53}, 2178, 1996)
\end{center}

\begin{abstract}
\noindent We investigate the scattering of an electron by an infinitely thin
and infinitely long straight magnetic flux tube in the framework of QED. We
discuss the solutions of the Dirac and Maxwell fields in the related external
pure AB potential and evaluate matrix elements and differential probabilities
for the bremsstrahlung process. The dependence of the resulting cross section
on the energy, direction and polarization of the involved particles is analyzed.
In the low energy regime a surprising angular asymmetry is found which results
from the interaction of the electron's magnetic moment with the magnetic field.

\medskip

\end{abstract}

\vspace{1cm}

%%%%%%%%%%%%%%%%%%%%%%%%%%%%%%%%%%%%%%%%%%%%%%%%%%%%%%%%%%%%%%%%%%%%%
%%%%%%%%%%%%%%%%%%%%%%%%%%%%%%%%%%%%%%%%%%%%%%%%%%%%%%%%%%%%%%%%%%%%%

\section{Introduction}

In classical physics the behaviour of charged particles in external
electro-magnetic fields is completely determined by the electric and magnetic
field strengths, $F_{\mu\nu}=\partial_{\mu}A_{\nu}-\partial_{\nu}A_{\mu}$,
which fix their trajectories by way of the resulting forces. For quantum
particles, however, the concept of field strengths, which are gauge invariant
local quantities, is insufficient. This became evident with the prediction of
the Aharonov-Bohm (AB) effect \cite{Aharonov59} (see also \cite{Franz39,
Ehrenberg49}), which states the influence of magnetic fluxes on quantum
systems. It was shown that it is the line integrals of the vector potential
over closed paths which produce the additional effects on quantum systems
\cite{Yang75}. The path-dependent phase factors $\exp\left(ie\oint
A_{\mu}dx_{\mu}\right)$, which form a class of topological, and again, gauge
invariant quantities, produce phase shifts in wave functions. A number of
remarkable experiments was made to observe the resulting interference pattern
of an electron beam scattered by a thin solenoid. For a comprehensive review
see \cite{Olariu85, Peshkin89}. In solid state physics the manifestation of
the AB-effect brought new unexpected results \cite{Lee85, Bergmann84}. The
influence of magnetic fluxes on vacuum polarization of quantum fields was
discussed in \cite{Sereb86}. For a detailed investigation of the AB effect
and its physical consequences it is essential not only to consider the
scattering but also accompanying effects. Among them bremsstrahlung is
supposed to be the most significant one. It was discussed at first for
non-relativistic particles in the dipole approximation in \cite{Sereb8586},
where the energetic and angular distributions were calculated. The polarization
properties of the bremsstrahlung were discussed for relativistic spinless
particles in \cite{Gal'tsov90}. The aim of the present paper is a detailed
investigation of the bremsstrahlung emitted by a Dirac electron being scattered
in the AB potential. This process is evidently of greatest interest in view of
a possible experimental verification using polarizable electron beams.

As usual we will consider the idealized case of an infinitely thin, infinitely
long straight solenoid (magnetic string), lying along the $z$ axis. The vector
potential connected with this {\em pure AB} configuration reads
$$
{\vec A} = {\Phi\over 2\pi\rho} {\vec e}_\varphi\,,
$$
where $\Phi$ is the magnetic flux and $\rho$ the distance to the string. This
idealization raises the question of how to describe the electron wave function
near the string correctly. Mathematically this is related to the
non-selfadjointness of the Hamilton operator \cite{Hagen90} -\cite{pragm}. We will discuss this point in section 2, where we investigate the Dirac
equation in the presence of the pure AB potential. There we also find the exact
scattering solutions in terms of partial waves. The issue of spin changes
slightly the interpretation of the AB effect. As we will show the interaction
between spin and magnetic field leads to the Dirac wave functions which do not
vanish on the magnetic string and thus, in a way, a local element is added to
the non-locality of the AB effect.

In section 3 the matrix element for the bremsstrahlung process is calculated
and the differential radiative cross section is obtained. We analyze the
behaviour of the differential and total cross section at different energy
regimes and discuss its particular features characteristic for the Dirac
electron in sections 4 and 5.

\medskip

We use units such that $\hbar=c=1$ and take $e<0$ for the electron charge.

%%%%%%%%%%%%%%%%%%%%%%%%%%%%%%%%%%%%%%%%%%%%%%%%%%%%%%%%%%%%%%%%%%%%%
%%%%%%%%%%%%%%%%%%%%%%%%%%%%%%%%%%%%%%%%%%%%%%%%%%%%%%%%%%%%%%%%%%%%%

\section{Solutions of the Dirac and Maxwell field in the Aharonov-Bohm
potential}

The Dirac equation in an external magnetic field reads
\begin{equation}  \label{de}
i\partial_{t}\psi = H\psi,\quad H = \alpha_{i}(p_{i} - eA_{i}) + \beta M
\end{equation}
where $e$ is the electron charge. For the matrices $\alpha$ and $\beta$ we use
$$
\alpha_{i} = \pmatrix{      0    & \sigma_{i} \cr
                      \sigma_{i} &       0 \cr}, \quad
\beta = \pmatrix{      1   &  0\cr
                       0   &  -1 \cr}.$$
In cylindrical coordinates $(\rho,\;\varphi,\;z)$, the kinetic momenta are
given by
\begin{equation}
\pi_{\rho} = p_{\rho} = - i\partial_{\rho},\quad \pi_{\varphi}
= p_{\varphi} - eA_{\varphi}=-{i \over \rho}\partial_{\varphi}-eA_{\varphi}
\,,\quad
p_{3} = -i \partial_{z}
\end{equation}
and
$$
\sigma_{\rho} = \sigma_1 \cos{\varphi} + \sigma_2 \sin{\varphi}
\,, \quad
\sigma_{\varphi} = (-\sigma_1 \sin{\varphi} + \sigma_2 \cos{\varphi})
$$
where $\sigma_i$ are the Pauli matrices.

The vector potential for the {\em pure AB case} (magnetic string) has a nonzero
angular component  \cite{Aharonov59}
\begin{equation} \label{abp}
eA_{\varphi} = {e\Phi \over 2\pi\rho} = -{\Phi \over \Phi_{0}\rho}
= {\phi\over\rho}\,,
\end{equation}
($0\leq\rho<\infty$), where $\Phi$ is magnetic flux and $\Phi_{0}=2\pi /|e|$
is the magnetic flux quantum. It corresponds to a magnetic field along the
$z$ axis
\begin{equation}
B_z = {2\phi\over e\rho} \delta(\rho) \,,
\end{equation}
which points to the positive (negative) $z$ direction for $\phi<0$ ($\phi>0$).
It is the fractional part $\delta$ of the magnetic flux $\phi = N+\delta$,
$0<\delta<1$ which produces all physical effects. Its integral part $N$ will
appear as a phase factor $\exp(iN\varphi)$ in the solutions to the Dirac
equation.

The exact solution of the Dirac equation used for the discussion of the
scattering problem in the external Aharonov-Bohm field can be written in an
integral form as it was done for the Schr\"odinger equation in the original
paper by Y.~Aharonov and D.~Bohm \cite{Aharonov59}. For further calculations,
however, cylindrical modes are more convenient.

For the Dirac equation in the AB field the complete set of commuting operators 
is
\begin{eqnarray}
\hat{H}\,,\quad
\hat{p_3} := -i\partial_z\,,\quad
\hat{J_3} := -i\partial_{\varphi}+{1 \over 2} \Sigma_3 \,,\quad 
\hat{S_3} := \beta\Sigma_3+\gamma \; {p_3\over M}\,
\end{eqnarray}
where $\gamma := \pmatrix{0 & 1 \cr
                   1 & 0 \cr}$.
The corresponding eigenvalue equations are given by
\begin{eqnarray} \label{co}
\hat{H}\psi &=& E\psi\,,\\
\hat{p_3}\psi &=& p_3\psi\,,\\
\hat{J_3}\psi &=& j_3 \psi\,,\\
\label{s3}
\hat{S_3}\psi &=& s\psi\,,
\end{eqnarray}
where $E = \sqrt{p_\perp^2 + p_3^2 +M^2}$ is the energy, $p_3$ and $j_3$
are the $z$-components of linear and total angular momentum respectively;
$p_\perp$ denotes the radial momentum.
The eigenvalue of $\hat{J}_3$ is half-integer and we rewrite it by introducing
$l$, $j_3 =: l + N +1/2$. Here $l$ is an integer number and N is
fixed as above.
Note that $l+N$ denotes an integral part of the eigenvalue of $\hat{J_3}$ in 
contrast to the usual convention. The corresponding separation of a factor 
$\exp(i N \varphi)$ in the solutions of the Dirac equation will turn out to 
be convenient in the following calculations.
We introduced in eq.~(\ref{s3}) the operator $\hat{S}_3$ and not the helicity 
operator $\hat{S}_t = \Sigma_i (p_i -e A_i) / p$ which is often used.
Both of these operators commute in 
the relativistic case with the operators $\hat{H}$, $\hat{p}_3$ and 
$\hat{J}_3$, when a magnetic field of a fixed direction is present. 
This can be seen, for example, in \cite{Sokolov68}. We prefer to use $\hat{S}_3$ because in the nonrelativistic limit, which will be treated below, it describes the spin projection along the direction of the magnetic field. Its eigenvalue is given by $s=\pm \sqrt{1+p_3^2/ M^2}$. Solving these eigenvalue equations leads to a radial solution of Bessel type. 

As independent solutions we choose Bessel functions of the first kind with
positive and negative orders. Then the normalization condition for the partial
modes with quantum numbers $j = (p_{\perp},p_3,l,s),$
\begin{equation} \label{nc}
\int dx \psi^{\dagger}(j,x) \psi(j^{\prime},x) = \delta_{j, j^{ \prime}}
= {\delta_{s, s^{ \prime}} \delta_{l, l^{ \prime}}}
{ \delta(p_{3} - p_{3^{ \prime}})} {\delta(p_{ \perp} - p_{ \perp}^{ \prime})
\over  \sqrt{p_{\perp} p_{ \perp}^{ \prime}}} \,,
\end{equation}
fixes the solutions (for electron states with $E>0$) only for values of $l$
outside the interval $-1 < l-\delta <0$. Unless $l=0$ the Bessel functions of
negative order are not square integrable and therefore the normalized solutions
contain only the regular Bessel functions of positive order. One finds, for
$l\neq 0$, for the total mode function
\begin{equation} \label{ds}
\psi_{e}(j,x) = {1\over 2\pi}N_e \; e^{-iE_p t + ip_3z} e^{iN\varphi}
e^{i{\pi\over 2}|l|}
\pmatrix{ u \cr v\cr},
\end{equation}
where
\begin{eqnarray} \label{u}
u &=&
{1\over \sqrt{2s}}
\left(\begin{array}{c}
\displaystyle
\sqrt{E_p + sM}\sqrt{s+1}\; J_{\nu_1}(p_{\perp}\rho)
e^{ il\varphi}\\
\displaystyle
i\epsilon_3\epsilon_{l}\sqrt{E_p - sM}\sqrt{s-1}\; J_{\nu_2}(p_{\perp}\rho)
e^{i(l +1)\varphi}
\end{array}\right) \,,\\
\label{v}
v &=&
{1\over \sqrt{2s}}
\left(\begin{array}{c}
\displaystyle
\epsilon_3 \sqrt{E_p + sM}\sqrt{s-1} \;J_{\nu_1}(p_{\perp}\rho)
e^{il\varphi}\\
\displaystyle
i\epsilon_{l}\sqrt{E_p - sM}\sqrt{s+1} \; J_{\nu_2}(p_{\perp}\rho)
e^{i(l +1)\varphi}
\end{array}\right)\,,
\end{eqnarray}
and
\begin{equation}  \label{N}
N_e := {1\over\sqrt{2E_p}}\,,\quad
p_{\perp} := \sqrt{p^2 - p^2_3} = \sqrt{E_p^2 - M^2 - p^2_3}\,,\quad
s = \pm \sqrt{1+{p_3^2\over M^2}}\,,\quad
\epsilon_3 :={\rm sign} (s p_3),
\end{equation}
\begin{equation} \label{order}
\nu_1 := \left\{\begin{array}{r}
l-\delta \\
-l+\delta \end{array}\right.\,, \quad
\nu_2 := \left\{\begin{array}{r}
l+1-\delta \\
-l-1+\delta \end{array}\right. \,, \quad
\epsilon_l := \left\{\begin{array}{rl}
1 &\quad {\rm if}\ l \geq 0\\
-1 &\quad {\rm if}\ l<0
\end{array}\right.\,.
\end{equation}
In (\ref{order}) we already included without proof the case $l=0$. This
critical mode, however, requires a separate discussion, to which we will turn
now.

The solution for $l=0$ contains Bessel functions of positive and negative
orders. It can be written as
\begin{eqnarray} \label{sds}
u &=&
{N_0 \over \sqrt{2s}}
\left(\begin{array}{c}
\sqrt{E_p + sM}\sqrt{s+1}\;
[\sin\mu J_{-\delta}(p_{\perp}\rho)+\cos\mu J_{\delta}(p_{\perp}\rho)]\\
i\epsilon_3\sqrt{E_p - sM}\sqrt{s-1}\;
[\sin\mu J_{1-\delta}(p_{\perp}\rho) -\cos\mu J_{-1+\delta}(p_{\perp}\rho)]
e^{i\varphi}
\end{array}\right)\,, \\
v &=&
{N_0 \over \sqrt{2s}}
\left(\begin{array}{c}
\epsilon_3 \sqrt{E_p + sM}\sqrt{s-1} \;
[\sin\mu J_{-\delta}(p_{\perp}\rho)+\cos\mu J_{\delta}(p_{\perp}\rho)]\\
i\sqrt{E_p - sM}\sqrt{s+1} \;
[\sin\mu J_{1-\delta}(p_{\perp}\rho) -\cos\mu J_{-1+\delta}(p_{\perp}\rho)]
e^{i\varphi}
\end{array}\right)\,,
\end{eqnarray}
where $\mu$ is an arbitrary parameter. It is normalizable for any value of
$\mu$. Moreover one can see that there is no solution that behaves regular
at $\rho=0$ as it is the case for the spinless case. At least one component
of each two-spinor diverges at $\rho=0$. This is an obvious consequence of the
additional interaction between the magnetic moment of the electron and magnetic
field inside the solenoid. The question remains, what value of $\mu$ is to be
chosen?

Note that the first spinor component is square integrable but singular
for $l=0$ as well as for $l=1$. (One has to replace $\delta$ with $-l+\delta$
in eq.\ (\ref{sds}) in order to get the general solution.) But for $l=1$ 
the second component turns out to be not square integrable.
This is the reason why only one critical mode appears, in contrast to 
\cite{Hagen91}.

The zero mode problem is not specific for the Dirac equation. It is related to
the fact that the Hamilton operator is not selfadjoint in the presence of the
pure Aharonov-Bohm potential for any charged quantum system. This would cause
many problems for the unitary evolution of quantum systems unless  selfadjoint
extensions of these Hamilton operators exist. The related self-adjoint
extension procedure \cite{ReedSimon} permits to fix the parameter $\mu$ up to
an arbitrary constant. We shall not describe this method here. In \cite{pragm}
we presented an alternative method of treating the self-adjointness problem.
It essentially amounts to finding orthogonal states with respect to the radial
momentum $p_\perp$ for the critical mode ($l=0$). It then immediately leads to
the same condition on the parameter $\mu$ as the self-adjoint extension method
does:
\begin{equation} \label{tan}
\tan\mu = \alpha \left({p_{\perp}\over M}\right)^{2\delta}{M\over E_p+sM}\,,
\quad \alpha = {\rm const}\,, \quad
N_0={1\over\sqrt{1+\sin2\mu\cos\pi\delta}}\,.
\end{equation}

If these conditions are fulfilled, then the cylindrical modes to the Dirac
equation form a complete set of functions for any $\alpha$. Similar results
can be obtained for other quantum systems in the Aharonov-Bohm potential both
relativistic and nonrelativistic ones.  Note that the self-adjoint extension
method does not fix the open parameter $\alpha$ which determines the behaviour
of the wave function at the origin.  This situation is not satisfactory from
the physical point of view (in addition the calculations in the following
sections cannot be performed for arbitrary $\alpha$). We solve this problem as
follows: The pure Aharonov-Bohm potential is the limiting case of nonsingular
vector potentials of  real sole\-noids of finite radii. Hamiltonians in the
presence of realistic magnetic fields are selfadjoint operators, and one would
expect that the limiting procedure permits us to fix the arbitrary parameter
$\alpha$ completely. This was shown in the paper \cite{Hagen90} in general
and we will confirm this result for the special case of a cylindrical solenoid
with uniform magnetic field inside. The vector potential outside the solenoid
coincides with (\ref{abp}). Inside it depends on the distribution of the
magnetic field. We choose a uniform distribution, so that
\begin{equation} \label{abpi}
eA_{\varphi} = {\phi \rho \over a^2}\,, \quad \rm {if} \quad \rho < a\,,
\end{equation}
where $a$ is the radius of the solenoid with the magnetic field
$H=2\phi/ea^2$.

Solutions to the Dirac equation contain the Bessel functions with positive and
negative orders for $\rho>a$ and a confluent hypergeometric function for
$\rho<a$. The matching and normalization conditions will fix all arbitrary
coefficients in these solutions. It is evident that for $l \neq 0$ the pure
AB solution will be restored in the limiting case of vanishing radius. For
$l=0$ one can expect that this procedure will lead to a self-adjoint extension
of the pure AB Dirac operator. So we need to consider only this special case.

The equations for radial functions are
\begin{equation} \label{dei}
R''_{1,2}+{1\over \rho}R'_{1,2}-{1\over \rho^2}\left(l_{1,2}
-{\phi \rho^2\over a^2} \right)^2R_{1,2}+p^2_{1,2}R_{1,2}=0 \,,
\end{equation}
with
$$
l_{1,2} = \pmatrix{      N   \cr
                        N+1  \cr}, \quad
p^2_{1,2}=p^2_{\perp} \pm {2\phi \over a^2.}$$
Solutions to these equations that are regular at $\rho =0$ are
\begin{equation} \label{dsi}
R_{1,2}=x^{{l_{1,2}\over 2}}
e^{-{x\over 2}} \Phi(A_{1,2}, C_{1,2}; x)
\end{equation}
where $\Phi(A, C ;x)$ is the confluent hypergeometric function with
$$
x=|\phi|\;{\rho^2\over a^2}, \quad A_{1,2}= -{p^2_{1,2}a^2 \over 4|\phi|}
+{1\over 2}, \quad C_{1,2}=l_{1,2}+{1\over 2}\,.
$$
One can see from the matching conditions that external solution will not
contain the Bessel function with a negative order unless the parameter
$A_{1,2}$ goes to zero at $a \rightarrow 0.$ It happens for the $R_{1}$
component at $\phi>0$ when the magnetic field directs down and for $R_{2}$
component at $\phi<0.$ In these cases the interaction of the electron magnetic
moment with the magnetic field is attractive and the probability to find the
electron near the string increases.

Thus, at $N>0$ when the magnetic field points to the negative $z$ direction,
which we shall assume from now on, the upper two-spinor component
(corresponding to spin up) is enhanced and one needs to put the parameter
$\mu$ in the equation (\ref{tan}) and the solution (\ref{sds}) equal to
$\pi /2$. The self-adjoint extension parameter becomes $\alpha=\infty$.
Then each component of the zero mode solution (\ref{sds}) contains only one
Bessel function and therefore we can include the $l=0$ case in (\ref{order}),
what was to be shown. Eqn.(\ref{ds}) with (\ref{u}) - (\ref{order}) represents
the central result of this section.

Let us remark that the method of taking the zero radius limit of
a finite flux tube, as sketched above, does not give physically meaningful
results if an additional Coulomb potential is present \cite{HagenPark}.
However, we do not consider that case here.

\medskip

The expression (\ref{ds}) with (\ref{u}) and (\ref{v}) presents the partial
electron wave functions in terms of cylindrical modes. The {\em electron
scattering wave function} is obtained by a superposition
\begin{equation} \label{eswf}
\Psi ({\vec p}, s, x) := \sum_l c_l \psi_e (j, x) \,.
\end{equation}
With the coefficients
\begin{equation} \label{pcoef1}
c^{(+)}_l := e^{-il\varphi_p}\;e^{i{\pi\over 2}\epsilon_l \delta}
\end{equation}
this solution behaves at large distances like a plane wave, propagating in the
direction of $\vec{p}$ given by $p_x = p_\perp \cos \varphi_p$, $p_y = p_\perp
\sin\varphi_p$, $p_z=p_3$, plus an outgoing cylindrical wave,
$$
\Psi ({\vec p}, s, x) \sim
\left( e^{i\vec{p} \vec{x}}
+ f(\varphi) \sqrt{{i\over\rho}}e^{i p_\perp\rho + i p_3 z} \right)
e^{-i \omega t}
$$
Because of the damping of the cylindrical wave at large distances we may use
this superposition instead of plane waves. Accordingly, the coefficients
\begin{equation} \label{pcoef2}
c^{(-)}_l := e^{-il\varphi_p}\;e^{-i{\pi\over 2}\epsilon_l \delta}
\end{equation}
are used to form a scattering wave solution for an outgoing electron with an
ingoing cylindrical wave. With this different choice of the coefficients for
ingoing and outgoing electrons the interaction with the external AB-field will
be described correctly.

\medskip

The external Aharonov-Bohm field does not influence the {\em photon wave
function}. In cylindrical coordinates it reads
\begin{equation} \label{vp}
A_{\mu}^{\lambda}(\vec{k}, x) = {e_{\mu}^{(\lambda)}\over\sqrt{2\omega_k}}
e^{-i\omega_k t + ik_3 z} e^{ik_{\perp}\rho \cos(\varphi-\varphi_k)}
\end{equation}
where the polarization vectors for two physical transversal photons (in
cartesian coordinates)
\begin{equation} \label{polv}
e^{(\sigma)} := (0,\; -\sin \varphi_k, \;\cos \varphi_k, \; 0), \quad
e^{(\pi)} := {1\over \omega_k}(0,\; -k_3\cos \varphi_k,\; -k_3\sin \varphi_k,
\; k_{\perp})
\end{equation}
correspond to two linear polarization states.

%%%%%%%%%%%%%%%%%%%%%%%%%%%%%%%%%%%%%%%%%%%%%%%%%%%%%%%%%%%%%%%%%%%%%
%%%%%%%%%%%%%%%%%%%%%%%%%%%%%%%%%%%%%%%%%%%%%%%%%%%%%%%%%%%%%%%%%%%%%

\section{Matrix elements and differential cross sections for the process
$e \rightarrow e + \gamma$}

Cross sections are usually related to plane wave states, and in our case to the
scattering states (\ref{eswf}). The cylindrical modes (\ref{ds}) have a
vanishing radial flux and therefore do not describe ingoing or outgoing
particles. They are, however, convenient for calculating matrix elements and
we shall use these matrix elements as starting point for the calculation of the
cross section which refers to scattering states.

%%%%%%%%%%%%%%%%%%%%%%%%%%%%%%%%%%%%%%%%%%%%%%%%%%%%%%%%%%%%%%%%%%%%%
\subsection{Matrix elements for cylindrical modes}

The matrix element for the bremsstrahlung process for an ingoing electron with
quantum numbers $j_p = (p_{\perp}, p_3,l,s)$ leading to an outgoing electron
with quantum numbers $j_q = (q_{\perp}, q_3,n,r)$ and a photon with quantum
numbers $j_k=(\vec{k},\; \lambda)$ for physical states $\lambda = (\sigma,\;
\pi)$ has the usual form
\begin{equation} \label{me1}
\widetilde M_{\lambda}(j_p, j_q) = -i \, \langle j_q, j_k|S^{(1)}|j_p \rangle
= -e \int {d^4x}\bar\psi_{e}(j_{q},x) \;{A^{\ast}}^{\lambda}_{\mu}(\vec{k},x)
\gamma^{\mu}\; \psi_{e}(j_{p},x)
\end{equation}
whereby gamma matrices are written in terms of Pauli matrices as
\begin{equation}   \label{gamma}
\gamma_i = \pmatrix{
         0    & \sigma_i \cr
    - \sigma_i &       0 \cr},
\end{equation}
so that
\begin{equation}  \label{alphas}
{e^{\ast}}_{\mu}^{(\lambda)} \gamma_{\mu} = \pmatrix{
        0    &  \alpha_{\lambda} \cr
      -\alpha_{\lambda}  &  0  \cr}, \;
\alpha_{\sigma} = \pmatrix{
        0    & -i e^{-i\varphi_k} \cr
        i e^{i\varphi_k}&  0  \cr}, \;
\alpha_{\pi} = \pmatrix{
       k_{\perp}   & -k_3 e^{-i\varphi_k} \cr
     -k_3 e^{i\varphi_k} &  - k_{\perp} \cr}{1\over\omega_k}.
\end{equation}
Using the expressions (\ref{ds}), (\ref{u}), (\ref{v}) and (\ref{vp}), we can
rewrite the matrix element (\ref{me1}) in the form
\begin{equation} \label{me2}
\widetilde M_{\lambda}(j_p, j_q)
=-e{1\over{2\sqrt{2\omega_{k}E_{q}E_{p}}}} e^{(i{\pi\over 2}(|l|-|n|)}
\delta(E_p-E_q-\omega_{k})\delta(p_3-q_3-k_3)\;m_{\lambda} \,,
\end{equation}
with
\begin{eqnarray} \label{me3}
m_{\lambda} &:=& \int\rho d\rho d\varphi
e^{-ik_{\perp}\rho\cos(\varphi-\varphi_k)} \;
\left[u^{\dagger}(q)\alpha_{\lambda}v(p)
+ v^{\dagger}(q)\alpha_{\lambda}u(p)\right ] \nonumber\\
&=& e^{i(l-n)\varphi_k}\;\int\rho d\rho d\varphi
e^{-ik_{\perp}\rho\cos(\varphi-\varphi_k)} K_{\lambda}(\rho, \varphi)\,.
\end{eqnarray}
For the polarization state $\lambda = \sigma$ we have
\begin{eqnarray} \label{Ks}
K_{\sigma}(\rho, \varphi) &=&
R\left[\epsilon_l \sqrt{E_q+rM}\sqrt{E_p-sM} J_{\nu_1}(q_{\perp}\rho)
J_{\nu_2}(p_{\perp}\rho) e^{i(l-n+1)(\varphi-\varphi_k)}\right. \nonumber\\
&&+ \left.\epsilon_n \sqrt{E_q-rM}\sqrt{E_p+sM} J_{\nu_2}(q_{\perp}\rho)
J_{\nu_1}(p_{\perp}\rho) e^{i(l-n-1)(\varphi-\varphi_k)}\right],
\end{eqnarray}
and for the polarization state $\lambda = \pi$
\begin{eqnarray} \label{Kp}
K_{\pi}(\rho, \varphi) &=&
{k_{\perp}\over\omega_k}\;R_{\perp}\left[ \sqrt{E_q+rM}\sqrt{E_p+sM}
J_{\nu_1}(q_{\perp}\rho) J_{\nu_1}(p_{\perp}\rho) \right.\nonumber\\
&& \quad
\left. - \epsilon_n\epsilon_l \sqrt{E_q-rM}\sqrt{E_p-sM}
J_{\nu_2}(q_{\perp}\rho) J_{\nu_2}(p_{\perp}\rho) \right]
e^{i(l-n)(\varphi-\varphi_k)} \nonumber\\
&& - {ik_3\over\omega_k}\;R\left[ \epsilon_l \sqrt{E_q+rM}\sqrt{E_p-sM}
J_{\nu_1}(q_{\perp}\rho) J_{\nu_2}(p_{\perp}\rho)
e^{i(l-n+1)(\varphi-\varphi_k)} \right. \nonumber\\
&& \quad
\left. - \epsilon_n \sqrt{E_q-rM}\sqrt{E_p+sM} J_{\nu_2}(q_{\perp}\rho)
J_{\nu_1}(p_{\perp}\rho) e^{i(l-n-1)(\varphi-\varphi_k)}\right] \,,
\end{eqnarray}
where
\begin{eqnarray}
R &:=&
{\sqrt{r+1}\sqrt{s+1}+\epsilon_3(q)\epsilon_3(p)\sqrt{r-1}\sqrt{s-1} \over
2\sqrt{sr}}\,,\\
R_{\perp} &:=&
{\epsilon_3(q)\sqrt{r-1}\sqrt{s+1}+\epsilon_3(p)\sqrt{r+1}\sqrt{s-1} \over
2\sqrt{sr}}\,.
\end{eqnarray}
Integrating over $\varphi$ we obtain
\begin{eqnarray} \label{mes2}
m_{\sigma} &=&
2\pi e^{i(l-n)\varphi_k} R \nonumber\\
&&
\times\left[\epsilon_l \sqrt{E_q+rM}\sqrt{E_p-sM} e^{-i{\pi\over 2}(l-n+1)}
\int\rho d\rho J_{\nu_1}(q_{\perp}\rho) J_{\nu_2}(p_{\perp}\rho)
J_{l-n+1}(k_{\perp}\rho)  \right. \nonumber\\
&&
+ \left.
\epsilon_n \sqrt{E_q-rM}\sqrt{E_p+sM} e^{-i{\pi\over 2}(l-n-1)}\int\rho d\rho
J_{\nu_2}(q_{\perp}\rho) J_{\nu_1}(p_{\perp}\rho) J_{l-n-1}(k_{\perp}\rho)
\right].
\end{eqnarray}
and
\begin{eqnarray} \label{mep2}
m_{\pi} &=&
2\pi e^{i(l-n)\varphi_k}\;e^{-i{\pi\over 2}(l-n)} \nonumber\\
&&
\times \left\{{k_{\perp}\over\omega_k}\;R_{\perp}
\left[\sqrt{E_q+rM}\sqrt{E_p+sM} \int\rho d\rho J_{\nu_1}(q_{\perp}\rho)
J_{\nu_1}(p_{\perp}\rho)J_{l-n}(k_{\perp}\rho) \right.\right.\nonumber\\
&& \quad
\left.\left. - \epsilon_n\epsilon_l \sqrt{E_q-rM}\sqrt{E_p-sM} \int\rho d\rho
J_{\nu_2}(q_{\perp}\rho) J_{\nu_2}(p_{\perp}\rho) J_{l-n}(k_{\perp}\rho)\right]
\right. \\
&&
-\left. {k_3\over\omega_k}\;R\left[ \epsilon_l \sqrt{E_q+rM}\sqrt{E_p-sM}
\int\rho d\rho J_{\nu_1}(q_{\perp}\rho) J_{\nu_2}(p_{\perp}\rho)
J_{l-n+1}(k_{\perp}\rho) \right.\right.\nonumber\\
&& \quad
\left.\left. +\epsilon_n \sqrt{E_q-rM}\sqrt{E_p+sM} \int\rho d\rho
J_{\nu_2}(q_{\perp}\rho) J_{\nu_1}(p_{\perp}\rho) J_{l-n-1}(k_{\perp}\rho)
\right]\right\}\,.\nonumber
\end{eqnarray}

It follows from the energy conservation $\delta$ term in (\ref{me2}) that the
radial momenta of the particles fulfill $p_{\perp} > q_{\perp}+k_{\perp}$. The
excess of radial momentum, $\Delta = 1- (q_\perp + k_\perp)/p_\perp$, is
transmitted to the flux tube. For this case, using formulae [6.578(3),
6.522(14)] of \cite{Gradshteyn80}, one can see that the integrals over
$\rho$ vanish unless $(l+{1\over 2})(n+{1\over 2}) < 0$. This inequality is
fulfilled at $l\geq 0, n<0$ and $n\geq 0, l<0$, and the nonvanishing integrals
are of the type
\begin{eqnarray} \label{int}
J(\alpha, \beta) &:=& \int_{0}^{ \infty} \rho d\rho
J_{\alpha}(p_{\perp}\rho\sin{A}\cos{B})
J_{\beta}(p_{\perp}\rho\cos{A}\sin{B}) J_{\beta-\alpha}(p_{\perp}\rho) \nonumber\\
&=&
{2\sin\pi\alpha\over \pi p_{\perp}^2
\cos(A+B)\cos(A-B)}\left({\sin{A} \over \cos{B}}
\right)^{\alpha}\left({\sin{B} \over \cos{A}}
\right)^{\beta} \,,
\end{eqnarray}
with $ q_{\perp} = p_{\perp}\sin{A}\cos{B}, \; k_{\perp} =
p_{\perp}\sin{B}\cos{A}$.

The partial wave analysis of the bremsstrahlung process in the Abahronov-Bohm 
potential shows a rather unexpected feature:
The process turns out to be forbidden unless the quantum numbers $l$ and $n$ 
of the ingoing and outgoing electron have opposite signs. 
This in turn implies that their kinetic angular momentum projections, 
$[\vec{r} \times (\vec{p}-e\vec{A})]_3 = - i \partial_\varphi -\phi$, have opposite signs.
To see this one calculates the corresponding expectation values and finds
\begin{eqnarray}
\langle [\vec{r} \times (\vec{p}-e\vec{A})]_3 \rangle &=& 
\langle - i \partial_\varphi -\phi \rangle \nonumber\\
&=& l + {1\over 2} - \delta - {1\over 2} s {M\over E_p} 
= l - \delta + {1\over 2} \left( 1 \pm {\sqrt{M^2 +p_3^2}\over\sqrt{M^2 +p_3^2 +p_\perp^2}}\right)\,,
\end{eqnarray}
from which it follows that the sign of $l$ equals the sign of 
$\langle [\vec{r} \times (\vec{p}-e\vec{A})]_3 \rangle$ for all values of 
$\delta$, $s$, and $l\neq0$.
In the framework of a semiclassical picture this leads to the interpretation
that the electrons need to pass the magnetic string in opposite directions.
Apparently this is necessary for the ingoing electron to give the excess of
its radial momentum to the string and to emit a real bremsstrahlung photon.

Using expression (\ref{int}) we find for the matrix element (\ref{mes2})
\begin{eqnarray} \label{mes3}
m_{\sigma} &=&
- 2\pi i e^{i(l-n)\varphi_k} e^{-i{\pi\over 2}|l-n|}R \nonumber\\
&&
\times\left\{\Theta(l\geq 0)\Theta(n<0)\left[\sqrt{E_q+rM}\sqrt{E_p-sM}\;
J(-n+\delta, l-n+1) \right.\right.\nonumber\\
&&
\quad +\left.\left.\sqrt{E_q-rM}\sqrt{E_p+sM}\;J(-n-1+\delta, l-n-1) \right]
\right.\nonumber\\
&&
+\left.\Theta(l<0)\Theta(n\geq 0)\left[\sqrt{E_q+rM}\sqrt{E_p-sM}\;
J(n-\delta, -l+n-1) \right.\right. \nonumber\\
&&
\quad+ \left.\left. \sqrt{E_q-rM}\sqrt{E_p+sM}\;J(n+1-\delta, -l+n+1)
\right]\right\}\nonumber\\
&=&
- {4i R\; e^{i(l-n)\varphi_k -i{\pi\over 2}|l-n|+i\pi n}\; \sin\pi\delta\over
\sqrt{p_{\perp}^4-2p_{\perp}^2(q_{\perp}^2+k_{\perp}^2)
+(q_{\perp}^2-k_{\perp}^2)^2}}(ab)^{|n|}b^{|l|} \\
&&
\times\left\{\sqrt{E_q+rM}\sqrt{E_p-sM}\left[\Theta(l\geq 0)\Theta(n<0)
a^{\delta}b - \Theta(l<0)\Theta(n\geq 0) a^{-\delta}b^{-1}\right]\right.\nonumber\\
&&
\left.-\sqrt{E_q-rM}\sqrt{E_p+sM}\left[\Theta(l\geq 0)\Theta(n<0)
a^{-1+\delta}b^{-1} - \Theta(l<0)\Theta(n\geq 0) a^{1-\delta}b\right]\right\}
\,,\nonumber
\end{eqnarray}
where we have introduced
\begin{eqnarray}
a &:=&
{\sin{A} \over \cos{B}} = {p_{\perp}^2-k_{\perp}^2+q_{\perp}^2
- \sqrt{p_{\perp}^4-2p_{\perp}^2(q_{\perp}^2+k_{\perp}^2)
+(q_{\perp}^2-k_{\perp}^2)^2}\over 2p_{\perp} q_{\perp}} \,, \\
b &:=&
{\sin{B} \over \cos{A}} = {p_{\perp}^2+k_{\perp}^2-q_{\perp}^2
- \sqrt{p_{\perp}^4-2p_{\perp}^2(q_{\perp}^2+k_{\perp}^2)
+(q_{\perp}^2-k_{\perp}^2)^2}\over 2p_{\perp} q_{\perp}}\,.
\end{eqnarray}
In the same way we find for the matrix element (\ref{mep2})
\begin{equation}
m_{\pi} = 2\pi e^{i(l-n)\varphi_k}\;e^{-i{\pi\over 2}|l-n|}
\left[\Theta(l\geq 0) \Theta(n<0)\;L + \Theta(l<0)\Theta(n\geq 0)\;N \right]
{1\over\omega_k}\,,
\end{equation}
with
\begin{eqnarray}
L &:=&
k_{\perp}R_{\perp}\left[\sqrt{E_q+rM}\sqrt{E_p+sM}\; J(-n+\delta, l-n)
\right.\nonumber\\
&&
\quad + \left. \sqrt{E_q-rM}\sqrt{E_p-sM}\;J(-n-1+\delta, l-n) \right] \\
&&
-k_3 R \left[\sqrt{E_q+rM}\sqrt{E_p-sM}\; J(-n+\delta, l-n+1) \right. \nonumber\\
&&
\quad - \left. \sqrt{E_q-rM}\sqrt{E_p+sM}\; J(-n-1+\delta, l-n-1)\right]\,,\nonumber
\end{eqnarray}
and
\begin{eqnarray}
N &:=&
k_{\perp}R_{\perp}\left[\sqrt{E_q+rM}\sqrt{E_p+sM}\; J(n-\delta, n-l)
\right. \nonumber\\
&&
\quad + \left. \sqrt{E_q-rM}\sqrt{E_p-sM}\;J(n+1-\delta, n-l) \right]  \\
&&
- k_3 R \left[ \sqrt{E_q+rM}\sqrt{E_p-sM}\; J(n-\delta, n-l-1)\right. \nonumber\\
&&
\quad - \left. \sqrt{E_q-rM}\sqrt{E_p+sM}\; J(n+1-\delta, n-l+1)\right]\,, \nonumber
\end{eqnarray}
so that
\begin{eqnarray} \label{mep3}
m_{\pi} &=&
{4 e^{i(l-n)\varphi_k -i{\pi\over 2}|l-n|+i\pi n}\; \sin\pi\delta\over
\sqrt{p_{\perp}^4-2p_{\perp}^2(q_{\perp}^2+k_{\perp}^2)
+(q_{\perp}^2-k_{\perp}^2)^2}} (ab)^{|n|}b^{|l|}\;{1\over\omega_k} \nonumber\\
&&
\times\left\{
k_{\perp} R_{\perp}\left[\sqrt{E_q+rM}\sqrt{E_p+sM}
\left( \Theta(l\geq 0)\Theta(n<0)a^{\delta}
-\Theta(l<0)\Theta(n\geq 0)a^{-\delta} \right)\right.\right.\nonumber\\
&&
\quad \left. -\sqrt{E_q-rM}\sqrt{E_p-sM}
\left( \Theta(l\geq 0)\Theta(n<0)a^{-1+\delta}
-\Theta(l<0)\Theta(n\geq 0) a^{1-\delta} \right)\right] \\
&&
- k_3 R \left[\sqrt{E_q+rM}\sqrt{E_p-sM}
\left(\Theta(l\geq 0) \Theta(n<0)a^{\delta}b
- \Theta(l<0)\Theta(n\geq 0)a^{-\delta}b^{-1} \right) \right. \nonumber\\
&&
\quad \left.\left. +\sqrt{E_q-rM}\sqrt{E_p+sM}
\left(\Theta(l\geq 0) \Theta(n<0)a^{-1+\delta}b^{-1}
- \Theta(l<0)\Theta(n\geq 0)a^{1-\delta}b \right)\right]\right\}. \nonumber
\end{eqnarray}
Eq.~(\ref{me2}) together with (\ref{mes3}) and (\ref{mep3}) is the final
expression for the bremsstrahlung matrix element with respect to cylindrical
modes.

%%%%%%%%%%%%%%%%%%%%%%%%%%%%%%%%%%%%%%%%%%%%%%%%%%%%%%%%%%%%%%%%%%%%%
\subsection{Cross section for scattering states}

It is now very easy to calculate the matrix element of the bremsstrahlung
process with respect to the electron scattering states (\ref{eswf}). Consider
an electron with momentum $\vec{p}$ and spin $s$ which is scattered by the
magnetic string and emits a photon with momentum $\vec{k}$ and polarization
$\lambda$. Denoting the momentum and spin of the scattered electron $\vec{q}$
and $r$ resp.~the corresponding matrix element reads
\begin{eqnarray} \label{pwma1}
M_{\lambda} &:=&
-i \, \langle (\vec{q},r),(\vec{k},\lambda)|S^{(1)}|(\vec{p},s)\rangle \nonumber\\
&=&
\sum_{l, n} c^{(+)}_l {c_{n}^{(-)}}^{\ast}
\widetilde M_{\lambda}(j_p, j_q) \nonumber\\
&=&
{-e\sqrt{2}\sin\pi\delta\over{\sqrt{\omega_{k}E_{q}E_{p}}}}\;
{\delta(E_p-E_q-\omega_{k})\; \delta(p_3-q_3-k_3) \over
\sqrt{p_{\perp}^4-2p_{\perp}^2(q_{\perp}^2+k_{\perp}^2)
+(q_{\perp}^2-k_{\perp}^2)^2}} \;\Sigma_{\lambda}
\end{eqnarray}
where the coefficients $c^{(+)}_l$ and $c^{(-)}_n$ are given in
eqs.~(\ref{pcoef1}) and (\ref{pcoef2}) and refer to the ingoing $(\vec{p},s)$
and outgoing electron $(\vec{q}, r)$ resp. It is
\begin{eqnarray}
i\Sigma_{\sigma} &:=&
\sum_{l,n}c_{l,n}\;(ab)^{|n|}\;b^{|l|}\;R \nonumber\\
&&
\times \left\{ \sqrt{E_q+rM}\sqrt{E_p-sM}
\left[\Theta(l\geq 0) \Theta(n<0) a^{\delta}b
- \Theta(l<0)\Theta(n\geq 0) a^{-\delta}b^{-1}\right]\right. \\
&&
\quad -\left.\sqrt{E_q-rM}\sqrt{E_p+sM}
\left[\Theta(l\geq 0)\Theta(n<0) a^{-1+\delta}b^{-1}
-\Theta(l<0)\Theta(n\geq 0) a^{1-\delta}b\right]\right\},\nonumber\\
\Sigma_{\pi} &:=&
\sum_{l,n}c_{l,n} \;(ab)^{|n|}\;b^{|l|}{1\over\omega_k} \nonumber\\
&&
\times \left\{k_{\perp}R_{\perp}
\left[\sqrt{E_q+rM}\sqrt{E_p+sM}\left( \Theta(l\geq 0) \Theta(n<0)a^{\delta}
-\Theta(l<0)\Theta(n\geq 0)a^{-\delta} \right)\right.\right. \\
&&
\quad -\left.\left. \sqrt{E_q-rM}\sqrt{E_p-sM}
\left( \Theta(l\geq 0)\Theta(n<0)a^{-1+\delta}
-\Theta(l<0)\Theta(n\geq 0) a^{1-\delta} \right)\right]\right. \nonumber\\
&&
- \left.k_3 R \left[\sqrt{E_q+rM}\sqrt{E_p-sM}
\left(\Theta(l\geq 0) \Theta(n<0)a^{\delta}b
- \Theta(l<0)\Theta(n\geq 0)a^{-\delta}b^{-1} \right) \right.\right. \nonumber\\
&&
\quad+ \left.\left. \sqrt{E_q-rM}\sqrt{E_p+sM}
\left(\Theta(l\geq 0) \Theta(n<0)a^{-1+\delta}b^{-1}
- \Theta(l<0)\Theta(n\geq 0)a^{1-\delta}b \right)\right]\right\}\,,\nonumber
\end{eqnarray}
with the new coefficients
$$
c_{l,n} := e^{-il(\varphi_p-\varphi_k)+in(\varphi_q-\varphi_k)
+i{\pi\over 2}(\epsilon_l + \epsilon_n)\delta }\,.
$$
After performing the sums over the angular quantum numbers $l, n$ and sorting
with respect to the flux parameter $\delta$ we obtain the following closed
expression.
\begin{equation} \label{sl}
\Sigma_{\lambda} =
\left(A_{\lambda}\;\Sigma_+ \;a^{\delta}
- B_{\lambda}\;\Sigma_-\;a^{-\delta}\right)\,,
\end{equation}
with
\begin{eqnarray} \label{sigma+}
\Sigma_+ &:=&
{1\over 1-ab\;e^{-i(\varphi_p-\varphi_k)}}
{b\;e^{-i(\varphi_q-\varphi_k)}\over 1-b\;e^{-i(\varphi_q-\varphi_k)}}\,, \\
\label{sigma-}
\Sigma_- &:=&
{ab\;e^{i(\varphi_p-\varphi_k)}\over 1-ab\;e^{i(\varphi_p-\varphi_k)}}
{1\over 1-b\;e^{i(\varphi_q-\varphi_k)}}\,,
\end{eqnarray}
\begin{eqnarray} \label{ABsi}
A_{\sigma} &:=&
R \left[ b\;\sqrt{E_q+rM}\sqrt{E_p-sM}-{1\over ab}\;\sqrt{E_q-rM}\sqrt{E_p+sM}
\right]\,, \\
B_{\sigma} &:=&
R \left[{1\over b}\;\sqrt{E_q+rM}\sqrt{E_p-sM} - ab\;\sqrt{E_q-rM}\sqrt{E_p+sM}
\right]\,, \\
\label{ABpi}
A_{\pi} &:=&
{k_{\perp}R_{\perp} \over \omega_k} \left[\sqrt{E_q+rM}\sqrt{E_p+sM}
- {1\over a}\;\sqrt{E_q-rM}\sqrt{E_p-sM}\right] \nonumber\\
&&
-{k_3 \over \omega_k} R \left[ (b\;\sqrt{E_q+rM}\sqrt{E_p-sM}
+ {1\over ab}\;\sqrt{E_q-rM}\sqrt{E_p+sM} \right]\,, \\
B_{\pi} &:=&
{k_{\perp} R_{\perp}\over \omega_k}
\left[\sqrt{E_q+rM}\sqrt{E_p+sM} - a\;\sqrt{E_q-rM}\sqrt{E_p-sM}\right] \nonumber\\
&&
- {k_3 \over \omega_k}R \left[{1\over b}\;\sqrt{E_q+rM}\sqrt{E_p-sM}
+ ab\;\sqrt{E_q-rM}\sqrt{E_p+sM} \right]\,.
\end{eqnarray}
Eqn.~(\ref{pwma1}) and (\ref{sl}) together with (\ref{sigma+})-(\ref{ABpi})
contain the results for the bremsstrahlung matrix element with respect to
scattering states.

\medskip

Based on the matrix element (\ref{pwma1}) we evaluate the differential
probability per unit length of string solenoid for the bremsstrahlung process:
\begin{equation} \label{dp}
dW_{\lambda} = W_{\lambda}\;q_{\perp}dq_{\perp} d\varphi_q \;
k_{\perp}dk_{\perp}d\varphi_k dq_3 dk_3 \,,
\end{equation}
where
\begin{equation} \label{Wl}
W_{\lambda} := \left| M_\lambda \right|^2
= {e^2\;\sin^2\pi\delta\over 8\pi^4 \omega_{k}E_{q}E_{p}} \cdot \;
\delta(E_p-E_q-\omega_{k})\; \delta(p_3-q_3-k_3)\;
P_{\lambda}(\vec{q}, \vec{k}) \,,
\end{equation}
with
\begin{equation} \label{pl}
P_{\lambda}(\vec{q}, \vec{k}) :=
{|\Sigma_{\lambda}|^2 \over p_{\perp}^4-2p_{\perp}^2(q_{\perp}^2+k_{\perp}^2)
+(q_{\perp}^2-k_{\perp}^2)^2}\,.
\end{equation}
In order to calculate the absolute value of (\ref{sl}) we shall make use of
the fact that the Dirac equation (\ref{de}) in the external Aharonov-Bohm
field is covariant under boost transformation along the string direction.
This means that it is sufficient to treat the case of normal incidence of
the ingoing electron on the magnetic string and therefore we may perform all
calculations in the coordinate system in which $p_3=0$. No information will
be lost but the calculations become more simple in this case. For $p_3=0$ we
use, instead of $q_{\perp}$ and $k_{\perp}$, more convenient variables: the
dimensionless photon energy $\omega$ and the angle $\vartheta_k$ between the
photon momentum and the string direction, defined by
\begin{equation} \label{ox}
\omega := {\omega_k\over p_{\perp}}
= {p_{\perp}^2-q_{\perp}^2+k_{\perp}^2\over 2p_{\perp} \sqrt{p_{\perp}^2+M^2}}
\,,\quad
x :=\sin\vartheta_k
={2k_{\perp}\sqrt{p_{\perp}^2+M^2}\over p_{\perp}^2-q_{\perp}^2+k_{\perp}^2}
\,.
\end{equation}
Referring to these variables we have $k_{\perp}=p_{\perp}\omega x, \;
q_{\perp}=p_{\perp}\sqrt{1-2{\omega\over v}+\omega^2x^2}, \; v $ being the
velocity of the ingoing electron, and
\begin{equation} \label{ab2}
a={1-{\omega\over v}-{\omega\over v}\sqrt{1-v^2x^2}
\over\sqrt{1-2{\omega\over v}+\omega^2x^2}}\,,\quad
b={1-\sqrt{1-v^2x^2}\over vx}\,.
\end{equation}
The variable $\omega$ ranges from $0$ to $\omega_{\rm max} = v\;
(1+\sqrt{1-v^2x^2})^{-1}$ which corresponds to the minimal value of
$q_{\perp}=0.$
Calculating
\begin{eqnarray}
|\Sigma_+|^2 &=& a\;\Sigma\,, \quad
|\Sigma_-|^2 = {1\over a}\;\Sigma \,,\\
\Sigma_+\Sigma_-^{\ast} + \Sigma_+^{\ast}\Sigma_-^{\ast} &=&
{2\over ab^2}\;\Sigma^2
\left\{[(1+b^2)\cos\varphi_{pk}-2b][(1+a^2 b^2)\cos\varphi_{qk}-2ab]
\right.\nonumber\\
&& \left. - (b^2-1)(a^2b^2-1)\sin\varphi_{pk}\sin\varphi_{qk}\right\} \,,
\end{eqnarray}
with
$$
\Sigma :=
{ab^2\over (1+b^2-2b\cos\varphi_{pk})(1+a^2b^2-2ab\cos\varphi_{qk})}, \quad
\varphi_{pk} := \varphi_p - \varphi_k \,,
$$
we obtain
\begin{equation} \label{P1}
P_{\lambda} = {P^{(+)}_{\lambda}\;a^{2\delta} + P^{(-)}_{\lambda}\;a^{-2\delta}
+ P^{(0)}_{\lambda} \over 8 E_p^2 \omega^2 (1-v^2x^2)
(1-vx\cos\varphi_{pk})\;
(1-\omega v x^2- vx\sqrt{1-2{\omega\over v}+\omega^2 x^2} \cos\varphi_{qk})}\,,
\end{equation}
where, again, we sorted with respect to the dependence on the flux parameter
$\delta$. We find for the polarization state $\sigma$
\begin{eqnarray} \label{P1+-}
P^{(\pm)}_{\sigma} &=&
{1\over 2rs}\left\{
(rs+1)(1-v^2x^2)(2-2{\omega\over v}+ \omega^2x^2)
-\omega^2(1-x^2)(2-v^2x^2-\omega v x^2)\right. \nonumber\\
&&
\left.\mp \sqrt{1-v^2x^2}(2-\omega v x^2)
[(r+s){M \omega\over p_{\perp}} + s{E_p\over M}\omega^2(1-x^2)]\right\} \,, \\
\label{P10}
P^{(0)}_{\sigma} &=&
{1\over rs}{q_{\perp}\over p_{\perp}} F(\varphi) \;
\left\{2(rs+1)(1-v^2x^2)+v^2\omega^2x^2(1-x^2)\right\}\,,
\end{eqnarray}
with
\begin{equation}
F(\varphi) :=
{(\cos\varphi_{pk}-vx)[p_{\perp}(1-\omega v x^2) \cos\varphi_{qk}-vx q_{\perp}]
- p_{\perp}(1-v^2 x^2) \sin\varphi_{pk} \sin\varphi_{qk} \over
(1-vx\cos\varphi_{pk})\;
[ p_{\perp}(1-\omega v x^2)- vxq_{\perp} \cos\varphi_{qk}]}\,,
\end{equation}
and for the polarization state $\pi$
\begin{eqnarray} \label{P2+-}
P^{(\pm)}_{\pi} &=&
{1\over 2rs}\left\{
(rs+1)[2(1-x^2)(1-{\omega\over v}) + \omega^2 x^2 (1-v^2x^2)]\right.\nonumber\\
&&
\mp \sqrt{1-v^2x^2}\;
[(r+s){M \omega\over p_{\perp}}(2-2x^2-\omega vx^2)
+ s{E_p\over M}\omega^2(1-x^2) (2-2v^2x^2-\omega vx^2) \nonumber\\
&&
\left. + s{M\over p_{\perp}} 2\omega^2 vx^4]
- \omega^2 [(1+x^2)(1-v^2x^2)
+(1-x^2)(1-\omega v x^2)]\right\}\,, \\
\label{P20}
P^{(0)}_{\pi} &=&
-{1\over rs}{p_{\perp}\over q_{\perp}} F(\varphi)\;(1-x^2)
\left\{ 2(rs+1) - \omega^2v^2x^2\right\}\,.
\end{eqnarray}
The result for $P_\lambda$, and thus for the differential probability
$dW_\lambda$ of eq.~(\ref{dp}), is given by eq.~(\ref{P1}) together with
(\ref{P1+-})-(\ref{P20}).

\medskip

In real experiments one observes the momentum and polarization distributions
of scattered electrons and bremsstrahlung photons. Complete information about
this is contained in the differential radiative-scattering cross section per
unit length of the solenoid
\begin{equation} \label{dcs}
{d\sigma_{\lambda}\over d\omega_k d\Omega_k d\varphi_q}
= {e^2\;\sin^2\pi\delta\over 8\pi^4 p_{\perp}}\,\omega_k \;
P_{\lambda}(\vec{q}, \vec{k})\big|_{E_q=E_p-\omega_k, q_3=p_3-k_3}\,,
\end{equation}
whereby $d\Omega_k = d\cos\vartheta_k\, d\varphi_k$.

The formula (\ref{dcs}) together with (\ref{P1})-(\ref{P20}) contains the
complete information about energy, angular and polarization distributions for
the bremsstrahlung process. It allows to analyse correlations between energy
and direction of the emitted radiation. We will now discuss the cross section
at different regimes of the incoming electron's energy.

%%%%%%%%%%%%%%%%%%%%%%%%%%%%%%%%%%%%%%%%%%%%%%%%%%%%%%%%%%%%%%%%%%%%%
%%%%%%%%%%%%%%%%%%%%%%%%%%%%%%%%%%%%%%%%%%%%%%%%%%%%%%%%%%%%%%%%%%%%%

\section{The differential cross sections at different energies}

For the soft bremsstrahlung process with small photon energy the infrared
singularity appears in the cross section in the limit $\omega \rightarrow 0$
as usual. This happens for any charged particle being scattered, and is not
specific neither for the Dirac electron nor the Aharonov-Bohm field. The reason
for this singularity is the failure of the perturbative expansion for soft
photon emission.

\medskip

For low electron energy ($v\ll 1$) we have $\omega\leq {v\over 2}, \; a^2
\approx 1-{\omega_k \over E_p-M}$  and $s,r = \pm 1.$ Then
\begin{equation}
P^{(\pm)}_{\sigma} = 2\Theta(sr)\left(1-{\omega\over v}\mp s{\omega\over v}
\right)\,,
\quad P^{(\pm)}_{\pi} = \cos^2\vartheta_k P^{(\pm)}_{\sigma}\,,
\end{equation}
\begin{equation}
P^{(0)}_{\sigma} = 2\Theta(sr) a \cos(\varphi_{pk} + \varphi_{qk})\,,\quad
P^{(0)}_{\pi} = -\cos^2\vartheta_k P^{(0)}_{\sigma}\,,
\end{equation}
and the differential cross section takes the form
\begin{equation}  \label{dcsle}
{d\sigma_{\lambda}\over d\omega_k d\Omega_k d\varphi_q}
= {e^2\;\sin^2\pi\delta \over 32\pi^4\;M\;\omega_k}\;v\;\Theta(sr)\; S^{(s)}
\pmatrix{1 \cr
\cos^2\vartheta_k \cr}\,, \quad {\rm for}\quad
\left\{\begin{array}{c}
\lambda=\sigma\\
\lambda=\pi
\end{array}\right. \,,
\end{equation}
with
\begin{equation}
S^{(s)} :=
\left(1-{\omega_k \over E_p-M}\right)^{1+s\delta}
+  \left(1-{\omega_k\over E_p-M}\right)^{-s\delta}
+ 2\epsilon_{\lambda} \left(1-{\omega_k\over E_p-M}\right)^{1\over2}
\cos(\varphi_{pk}+\varphi_{qk})\,,
\end{equation}
where $\epsilon_{\lambda}= 1$ for $\lambda=\sigma $ and $\epsilon_{\lambda}
= -1$ for $\lambda=\pi$. The cross section of the bremsstrahlung process is
rather small. It is proportional to the classical electron radius, $r_0 =
e^2/4\pi M,$ and the  velocity $v$ of the ingoing electron.

Let us note some particular features of the low energy bremsstrahlung process
for the Dirac electron which can be read off from eq.~(\ref{dcsle}).

\noindent (i) The electron spin projection is conserved at low electron
energies as well as for soft photon emission at arbitrary energies.

\noindent (ii) The appearance of the additional factor $\cos^2\vartheta_k$ for
the photon polarization state $\pi$ is typical for the angular distribution of
radiation from nonrelativistic particles. But the angular distributions of the
outgoing electron and emitted photon are not independent on the angles
$\varphi_q$ and $\varphi_k$ as it is the case for spinless particles. In
addition to the first two terms, which appear for spinless particles too, for
the Dirac electron an additional term occurs. This Dirac-specific term is
dependent on the direction of the radiation relative to the momenta of the
ingoing and outgoing electron. However, it does not depend on the field
strength of the magnetic string and therefore it seems to be obvious that it
arises due to the interaction of the electron's and the photon's spin.

\noindent (iii) After integration over the angle $\varphi_q$ of the outgoing
electron only the first two terms remain and one finds
$$
S^{(s)}=
2\pi \left\{
\left(1-{\omega_k \over E_p-M}\right)^{1+s\delta}
+  \left(1-{\omega_k\over E_p-M}\right)^{-s\delta}
\right\}\,.
$$
The cross section then coincides for $s=-1$ with the one for nonrelativistic
spinless particles \cite{Sereb86}. In this case it is invariant under the
transformation $\delta\to 1-\delta$, and consequently also under $\Phi\to
-\Phi$. For $s=+1$, however, the interaction between the spin and the magnetic
field leads to an attractive interaction between the spin and the magnetic
field and results in a modification of the cross section. For spinless
particles and for Dirac electrons with spin antiparallel to the magnetic
string the wavefunctions vanish at the origin. Consequently their cross
sections coincide at low energies. For an electron with antiparallel spin
projection, $s=+1$ in our case, the wave function is amplified near the
string and this leads to a
different cross section.

\noindent (iv) The photon emission with electron spin-flip takes place for
higher electron energies. As a result of the scattering process the electron
tends to direct its spin antiparallel to the magnetic field. It leads to a
small self-polarization effect for the electron beam in the same manner as it
happens for synchrotron radiation  \cite{Sokolov68}.

We now discuss the angular distribution in the case of high electron energy,
when $\gamma = 1/\sqrt{1-v^2} \gg 1.$ Due to the presence of the factors
\begin{equation}
(1-vx\cos\varphi_{pk})\sim E_p\omega_k-\vec{p}\vec{k}
\end{equation}
and
\begin{equation}
(1-\omega v x^2- vx\sqrt{1-2{\omega\over v}+\omega^2 x^2} \cos\varphi_{qk})
\sim E_q\omega_k-\vec{q}\vec{k}
\end{equation}
in the denominator of (\ref{P1}) the bremsstrahlung cross section has a sharp
maximum in the neighborhood of the direction of the ingoing electron, and the
radiation is concentrated within a narrow cone around this direction. This cone
has an angular aperture of order of magnitude $\sim 1/\gamma$. The same cone
also contains the momentum of the scattered electron. Outside the probability
of radiative transition decreases rapidly.

%%%%%%%%%%%%%%%%%%%%%%%%%%%%%%%%%%%%%%%%%%%%%%%%%%%%%%%%%%%%%%%%%%%%%
%%%%%%%%%%%%%%%%%%%%%%%%%%%%%%%%%%%%%%%%%%%%%%%%%%%%%%%%%%%%%%%%%%%%%

\section{The total cross section}

Let us now integrate over final states in order to find the total cross
section. After integrating over the angle $\varphi_{q}$ of the outgoing
electron the term with $P^{(0)}$ disappears and we obtain
\begin{equation} \label{tcs1}
{d\sigma_{\lambda}\over d\omega_k d\Omega_k} = {e^2\;\sin^2\pi\delta
\over 32\pi^3 \omega_k}{v\over E_p} {P^{(+)}_{\lambda}\;a^{2\delta}
+ P^{(-)}_{\lambda}\;a^{-2\delta}\over (1-v^2x^2)^{3\over 2}
(1-vx\cos\varphi_{pk})}\,.
\end{equation}
Summing over polarization states of the outgoing electron (indicated by a bar)
we find
\begin{equation} \label{tcs2}
{d\bar{\sigma}_{\lambda}\over d\omega_k d\Omega_k}
= {e^2\;\sin^2\pi\delta \over 32\pi^3\;\omega_k}{v\over E_p}
{\bar{P}^{(+)}_{\lambda}\;a^{2\delta}
+ \bar{P}^{(-)}_{\lambda}\;a^{-2\delta}\over (1-v^2x^2)^{3\over 2}
(1-vx\cos\varphi_{pk})}\,,
\end{equation}
with
\begin{equation} \label{tP1+-}
\bar{P}^{(\pm)}_{\sigma}
= (1-v^2x^2)(2-2{\omega\over v}+\omega^2x^2)
\mp s \sqrt{1-v^2x^2}\;(2-\omega v x^2) {M \omega\over p_{\perp}}\,,
\end{equation}
and
\begin{equation} \label{tP2+-}
\bar{P}^{(\pm)}_{\pi} =
2(1-x^2)(1-{\omega\over v}) + \omega^2 x^2 (1-v^2x^2)
\mp s \sqrt{1-v^2x^2}\;(2-2x^2-\omega vx^2){M \omega\over p_{\perp}}\,.
\end{equation}
The integration over the angle $\varphi_k$ of the emitted photon leads to an
additional factor $2\pi (1-v^2x^2)^{-{1\over 2}}$. We find as result
\begin{equation} \label{tcs}
{d\bar{\sigma}_{\lambda}\over d\omega_k d\cos\vartheta_k}
= {e^2\;\sin^2\pi\delta \over 16\pi^2 \;\omega_k}{v\over E_p}
{\bar{P}^{(+)}_{\lambda}\;a^{2\delta}
+ \bar{P}^{(-)}_{\lambda}\;a^{-2\delta} \over (1-v^2x^2)^2 }\,.
\end{equation}
The remaining integrals over $d\omega_k$ and $d\vartheta_k$ can not be found
analytically for arbitrary electron energies.

For low electron energies we have
\begin{equation} \label{tcsle}
{d\bar{\sigma}_{\sigma}\over d\omega_k}
= 3 {d\bar{\sigma}_{\pi}\over d\omega_k}
= {r_0\;\sin^2\pi\delta \over \pi} {v\over\omega_k}
\left[\left(1-{\omega_k \over E_p-M}\right)^{1+s\delta}
+  \left(1-{\omega_k\over E_p-M}\right)^{-s\delta}\right]\,.
\end{equation}
This result can also be obtained from (\ref{dcsle}) by integrating over
$\varphi_q$, $\varphi_k$ and $\vartheta_k$, i.~e.~by first making the
approximation of low energy and summing over final states afterwards.

For high electron energies, $\gamma=(1-v^2)^{-{1\over 2}}\gg 1,$ the main
contribution to the cross section arises from  values of $\vartheta_k
\sim {\pi\over 2},$ or $x\sim 1.$ In this case  $a\approx 1$ if the
bremsstrahlung photon is not too hard, $\omega < \omega_{\rm max},$ and we
obtain for the cross section
\begin{equation}  \label{tcshe1}
{d\bar{\sigma}_{\lambda}\over d\omega_k}
\sim {r_0\;\sin^2\pi\delta \over\omega_k}f_{\lambda}(\omega_k)\,,
\end{equation}
where $f_{\lambda}(\omega_k)$ is a function of the photon energy. It means
that the form of the energy bremsstrahlung spectrum is independent on the
ingoing electron energy and polarization state. For hard photons,
$\omega\sim\omega_{\rm max},$ we have $a^2 \sim
(\omega_{\rm max}-\omega)/2\sqrt{1-v^2x^2},$ and the cross section behaves
asymptotically as
\begin{equation} \label{tcshe2}
{d\bar{\sigma}_{\lambda}\over d\omega_k}
\sim {r_0\;\sin^2\pi\delta \over\omega_k}\;\gamma^{\delta}
\left(1-{\omega_k\over E_p}\right)^{\delta} (a\mp bs) \,,
\end{equation}
where $a$ and $b$ are coefficients and the signs of $\mp$ correspond to
$\lambda = \sigma$ and $\pi$ resp. It means that the fraction of hard photons
increases with the ingoing electron energy. i.e. the bremsstrahlung photon
spectrum becomes more hard. Also there is a correlation between the electron
spin states and the photon polarization.

%%%%%%%%%%%%%%%%%%%%%%%%%%%%%%%%%%%%%%%%%%%%%%%%%%%%%%%%%%%%%%%%%%%%%
%%%%%%%%%%%%%%%%%%%%%%%%%%%%%%%%%%%%%%%%%%%%%%%%%%%%%%%%%%%%%%%%%%%%%

\section{Conclusions}

We have analyzed the bremsstrahlung of an electron being scattered by a
magnetic string, which, among other quantum processes accompanying the AB
scattering, seems to be the most interesting and significant one.

In addition to the AB interaction, resulting from the non-integrable phase
factors, which all quantum particles suffer, spin particles interact with the
magnetic field via their magnetic moments. This strongly influences their
behaviour near the flux tube. In the idealized case of an infinitely thin
magnetic string their wave functions do not vanish on the string and the
non-locality of the AB effect is modified by a local interaction. This
interaction leads to a specific behaviour of the cross section.

We evaluated the differential radiative cross section, which contains complete
information about energy, angular and polarization distribution of the
participating particles, as well as the total cross section and analysed them
for different energy regimes. For low electron energies an angular asymmetry in
the plane perpendicular to the string occurs. This result may be the most
interesting one for possible experimental observations.

The bremsstrahlung allows for attempts of detailed experimental investigation
of the AB effect. Of course, the observation of the AB effect, which is done
by means of electron interference and electron holography \cite{Peshkin89}, is
not a simple task and the radiative AB effect for polarized electron beams
requires a careful discussion.

We draw attention to a remarkable feature which is characteristic for quantum
processes in the presence of the AB string both for spinless and for spin
particles. The processes happen if the ingoing and outgoing electrons have
angular momentum projection of opposite signs. In a sense the electron needs
to circle the AB string to emit the bremsstrahlung photon. In this case it
can transmit a part of its perpendicular momentum to the string. The relative
excess, $\Delta = 1-(q_{\perp}+k_{\perp})/p_{\perp}$, is ranged from its
minimal value $\Delta=0$ at $\omega_k=0$ to the maximal value $\Delta=1$ at
$\omega_k=p_{\perp}v/2$ when the bremsstrahlung photon takes away all kinetic
energy of the electron.

Finally we point out the analogy to the physics of an electron passing a
cosmic string \cite{Audretsch94}. In this case an additional term appears
which results from the spin connection. It corresponds to the vector potential
term in the AB case.

%%%%%%%%%%%%%%%%%%%%%%%%%%%%%%%%%%%%%%%%%%%%%%%%%%%%%%%%%%%%%%%%%%%%%
%%%%%%%%%%%%%%%%%%%%%%%%%%%%%%%%%%%%%%%%%%%%%%%%%%%%%%%%%%%%%%%%%%%%%

\section*{Acknowledgments}

V.~S.~thanks J.~Audretsch and the members of his group at the University of
Konstanz for hospitality, collaboration and many fruitful discussions. This
work was supported by the Deutsche Forschungsgemeinschaft.

%%%%%%%%%%%%%%%%%%%%%%%%%%%%%%%%%%%%%%%%%%%%%%%%%%%%%%%%%%%%%%%%%%%%%
%%%%%%%%%%%%%%%%%%%%%%%%%%%%%%%%%%%%%%%%%%%%%%%%%%%%%%%%%%%%%%%%%%%%%


\begin{thebibliography}{99}

\bibitem{Aharonov59}
Y.~Aharonov and D.~Bohm, Phys.~Rev.~{\bf 119}, 485 (1959).

\bibitem{Franz39}
W.~Franz,  Verh.~Deutsch.~Physik.~Ges.~{\bf 20}, 65 (1939).

\bibitem{Ehrenberg49}
W.~Ehrenberg and R.~E.~Siday,  Proc.~Phys.~Soc.~London B {\bf 62}, 8 (1949).

\bibitem{Yang75}
T.~T.~Wu and C.~N.~Yang,  Phys.~Rev.~D {\bf 12}, 3864 (1975).

\bibitem{Olariu85}
S.~Olariu and I.~I.~Popescu,  Rev.~Mod.~Phys.~{\bf 47}, 339 (1985).

\bibitem{Peshkin89}
M.~Peshkin and A.~Tonomura,$\;$ {\it The Aharonov-Bohm effect},
Springer-Verlag,  Berlin (1989).

\bibitem{Lee85}
P.~A.~Lee and T.~V.~Ramakrishnan,  Rev.~Mod.~Phys.~{\bf 57}, 287 (1985).

\bibitem{Bergmann84}
G.~Bergmann,  Phys.~Rep.~{\bf 107}, 1 (1984).

\bibitem{Sereb86}
E.~M.~Serebryany\u\i,  Theor.~Math.~Phys.~{\bf 64}, 846 (1986).

\bibitem{Sereb8586}
E.~M.~Serebryany\u\i\ and V.~D.~Skarzhinski\u\i,
Sov.~Phys.~- Leb.~Inst.~Rep.~{\bf 6}, 56 (1988),
and in {\em Proc.~P.~N.~Lebedev Phys.~Inst.~{\bf 197}} (in Russian),
ed.~by M.~A.~Markov, Nauka, Moscow (1989)


\bibitem{Gal'tsov90}
D.~V.~Gal'tsov and S.~A.~Voropaev, Sov.~J.~Nucl.~Phys.~{\bf 51}, 1811 (1990).

\bibitem{Hagen90}
C.~R.~Hagen,  Phys.~Rev.~Lett.~{\bf 64}, 503 (1990).

\bibitem{Hagen91}
C.~R.~Hagen,  Int.~J.~Mod.~Phys.~A {\bf 6}, 3119 (1991).

\bibitem{Gerbert89}
Ph.~S.~de Gerbert,  Phys.~Rev.~D {\bf 40}, 1346 (1989).

\bibitem{pragm}
J.~Audretsch, U.~Jasper and V.~D.~Skarzhinsky, {\em A pragmatic approach to
the problem of the self-adjoint extension of Hamilton operators with the
Aharonov-Bohm potential}, to appear in J.~Phys.~A: Math.~Gen.~(1995)

\bibitem{ReedSimon}
M.~Reed and B.~Simon, {\it Methods of Modern Mathematical Physics vol.~II,
Fourier Analysis, Self-Adjointness}, Academic Press, New York (1975).

\bibitem{Gradshteyn80}
I.~S.~Gradshteyn and I.~M.~Ryzhik, {\it Table of
integrals, series and products}, Academic Press (1980).

\bibitem{Sokolov68}
A.~A.~Sokolov and I.~M.~Ternov, {\it Synchrotron radiation},
Akademie-Verlag, Berlin (1968).

\bibitem{Audretsch94}
J.~Audretsch, U.~Jasper, and V.~Skarzhinsky,  Phys.~Rev.~D {\bf 49}, 6576
(1994).


\bibitem{HagenPark}
C.~R.~Hagen and D.~K.~Park, {\em Relativistic Aharonov-Bohm-Coulomb problem},
{\tt hep-th/9410225} (1994).

\end{thebibliography}
\end{document}